\documentclass[12pt,preprint,eqsecnum,aps]{revtex4} 
\usepackage{times}
\usepackage{graphicx}
\usepackage[fleqn]{amsmath}
\usepackage{amsfonts}
\usepackage{amssymb}
\usepackage{amsopn}
\usepackage{xspace}
\usepackage{array}
\usepackage{epsfig}
\begin{document}
\title{ Considerations for a cosmological extension of modified Newtonian dynamics connections to conformal gravity and Rindler force theories}

\author{ Ehoud Pazy }

\affiliation{Department of Physics, University of Connecticut, Storrs, Connecticut 06269, USA \footnote{On leave from Department of Physics, NRCN, P.O.B. 9001, Beer-Sheva 84190, Israel}}

\numberwithin{figure}{section}

\begin{abstract}
Modified Newtonian dynamics (MOND) can be obtained by modifying the entropic formulation of gravity, this is achieved by considering the quantum statistical nature of the degrees of freedom on the holographic screen. Through this  frame work, we find some constraints on a cosmological extension for MOND, with no additional auxiliary fields. The connections between MOND to conformal gravity and Rindler force gravity are examined. These two alternative gravity theories are subsequently considered as possible cosmological extensions of MOND.
 \end{abstract}
 
 \maketitle
 
\section{Introduction}
\label{sec:intro}
In constructing theories describing the universe, in accordance with the principle of Occam's razor,  the hope is to have theories with little as possible assumptions explaining as much as possible of the observed phenomena. Newtonian dynamics was extremely successful in this sense, describing all the local dynamics of the solar system, aside from the problem of Mercury's perihelion procession. It turned out that by replacing Newtonian gravity, general relativity (GR)  was able to explain  away the perihelion mystery as well as predicting gravitational lensing, gravitational redshift and gravitational time dilation. Combining GR with the Standard Model of particle physics, the theory obtained was able to model much more of the astronomical observations and via this combined theory, an explicit formulation for the evolution of the universe was developed. Some of the major successes of this model are in explaining, the relative abundances of isotopes in the universe, the cosmic background radiation and the large structure in the universe. However, it soon became apparent that not all observed astronomical observations could be explained using  GR combined with the standard model (see for example \cite{Starkman,Milgrom12}). To explain the notably flat rotation curves of galaxies as originally studied in \cite{Rubin}, dark matter (DM) was introduced as a further ingredient. Theoretically the need for DM also arose when estimates were made for calculating the age of large-scale structures observed in the universe, it turned out there was not enough gravity out there to have grown the observed large-scale structure. At first hand, DM  as a theory, seems to do well, by adding the single assumption of dark matter,  many of the problems GR and the standard model fail to explain are addressed. All the same, it should be noted that  the ability of DM to fit observational data requires tuning ,i.e., the amount of DM, which needs to be added is specific to each galaxy or cluster observed. Once tuning is required the theory is not based on a single assumption any more, it now requires initial conditions. The amount of DM added is not obtained by some basic physical argument  rather it is introduced to fit observations. One should pay close attention to warning given by Feynman in describing the requirements from a physical theory, "We do not want to proceed in a fashion that would allow us to change the details of the theory at every place that we find it in conflict with experiment..." \cite{Feynman}. Moreover the modeling of the universe by adding mass sources, case by case, seems to fail in explaining the regularities revealed in the observational data, such as the baryonic-Tully-Fisher (TF) relation \cite{TF}. In an attempt to solve the flat rotation curve problem, without the introduction of DM,  Milgrom introduced in 1983 a phenomenological theory, modified Newtonian dynamics (MOND) \cite{Milgrom83}, a theory which involves a single acceleration scale $a_0\simeq 10^{-10} m s^{-2}$ and a universal interpolating function $\tilde{\mu}$, which is a function of the ratio of the acceleration, $g$, to $a_0$ (for a recent review see \cite{MONDRev}). By requiring that the MOND interpolating function be linear in the acceleration for the low acceleration limit $g <<a_0$, MOND obtains an asymptotic velocity, which does not depend on the rotation radius, thus the asymptotic rotation curve is automatically flat and the baryonic TF relation is satisfied. MOND consequently explains away the velocity discrepancies in galactic systems with no need for DM. 

Although MOND was designed for galaxies it turns out it might be connected to cosmological evolution as well. A hint in this direction is given by the physical scale of the MOND acceleration scale, 
\begin{equation}
\label{eq:cos_a0}
2 \pi a_0 \approx c H_0,
\end{equation}
where $H_0$ is the Hubble constant and $c$ is the speed of light. This relation seems to indicate that MOND is a much more general model then what it was intended to be and therefore it should have a natural extension to a relativistic theory. Recently many attempts have been made  at constructing a relativistic theory, which in the classical limit reduces to MOND (see e.g. \cite{MONDRev}); however, most of these involve addition of extra fields. Adding more fields contradicts the logic of Occam's razor, the principle with which we started, for this reason the approach considered in the current paper is different. We consider an underlaying physical model for MOND based on the quantum statistical modified entropic gravity (QSMEG) formulation introduced in Ref. \cite{Pazy12} and further explored in Ref. \cite{Pazy13}. Based on QSMEG we will consider some the modification needed for MOND to become a cosmological theory.

We start by briefly reviewing the QSMEG formulation in Sec. \ref{sec:cosmology}, showing how the universe's acceleration can be obtained in terms of it, up to some caveats. In Sec. \ref{sec:consideration} we introduce a naive cosmological extension of MOND, analyze its problems, and discuss a possible solution. Employing thermodynamic scaling arguments, we obtain from QSMEG how the MOND acceleration scale should be modified in a cosmological theory. In Sec. \ref{sec:complications} we review a further complication regarding a cosmological extension of MOND and how these problems can be avoided if one considers the modification implied by QSMEG. Two alternative gravity theories are considered as possible cosmological extensions of MOND in Sec. \ref{sec:connections} and the conclusions are presented in Sec. \ref{sec:conclusions}.

\section{The universe's acceleration related to holographic  surface energy in QSMEG}
\label{sec:cosmology}

The idea of entropic gravity was put forward by Verlinde  \cite{Verlinde11} and independently by Padmanabhan (for a review see \cite{Padmanabhan10}). In Verlind's entropic formulation of gravity, gravity is defined by thermodynamic relations on the holographic screen. The basic building blocks and assumptions involved in constructing this entropic model of gravity are: associating a temperature to the holographic screen through the Unruh relation, associating an entropy to screen and connecting it with the number of degrees of freedom on the holographic screen. Then by assuming thermodynamic equipartition of energy, Newton's gravitational law is obtained. In Ref. \cite{Pazy12} it was shown that considering a change in this entropic formulation of gravity, in which one considers the quantum statistics of the underlying degrees of freedom on the holographic screen, instead of assuming equipartition, MOND is obtained rather than Newtonian dynamics; this is the basis for QSMEG. In Ref. \cite{Pazy13} QSMEG was further developed; two new thermodynamical expressions for the MOND interpolating function $\tilde{\mu}$ were derived and  extensions of MOND to de Sitter space were considered. 
 
\subsection{Introducing QSMEG}
\label{sec:qsmeg}

Apart from QSMEG, quite a few recent attempts have been made to obtain MOND through entropic gravity, the key idea behind all these attempts is that in the entropic gravity formulation, MOND results through the modification of the equipartition relation for the degrees of freedom on the holographic screen. In \cite{Li10} a Debye model was considered for the excitations on the holographic screen thus restricting the excited degrees of freedom at low temperatures, whereas \cite{Kiselev11} considered collective excitations on the holographic screen thus obtaining MOND. Another idea for restricting the distribution of the energy between excitations was proposed in \cite{Klinkhamer2011} , in which a minimal temperature for excitations, on the holographic screen, connected to $a_0$ was suggested.
In \cite{Neto} a nonhomogeneous cooling of the holographic screen was considered resulting from a phase transition occurring at a critical temperature, which was related to $a_0$. 
In contrast to these works QSMEG simply assumes that degrees of freedom on the holographic screen should be treated through the quantum-statistical-mechanics formalism as described below.

In constructing QSMEG, we start by considering a mass, $M$, surrounded by a spherical holographic screen of radius, $R$, and area, $A$. The information on the holographic screen is stored in bits such that the number of bits is given by  
\begin{equation}
\label{eq:Area}
N={A c^3 \over G \hbar},
\end{equation}
where $G$ is Newton's gravitational constant. We regard the bits on the holographic screen as degrees of freedom and consider them to be in a thermal equilibrium at a temperature, $T$. When the temperature is large enough, the thermal energy of the holographic degrees of freedom is given by the equipartition  relation, in which an energy of, $k_B T /2$, is associated with each degree of freedom, where $k_B$ is the Boltzman constant, which for convenience, we will from here on consider energy units such that $k_B=1$. However, once the temperature is lowered, one needs to modify the equipartition relation according to the quantum statistics of the holographic degrees of freedom,
\begin{equation}
\label{eq:energy}
E_{th}={1 \over 2}\tilde{\mu} N T,
\end{equation}
where the function $\tilde{\mu} (T)$, gives the ratio of the thermal energy per degree of freedom to its equipartition value. This function will be shown to correspond to the MOND interpolating function and in terms of QSMEG it can be calculated analytically \cite{Pazy12}, simply by calculating the thermal energy of a two-dimensional gas,
\begin{equation}
\label{eq:mu_analytical}
{\tilde \mu}({T \over T_0})=-{6 \over \pi^2}{T  \over T_0} Li_2(-e^{{\mu}/T})- {\pi^2 \over 6}{T_0 \over T},
\end{equation}
where $T_0$ is some energy scale separating the quantum statistical regime from the classical regime and $ \mu$ is the chemical potential, which  should not to be confused with $\tilde{\mu}$, the MOND interpolating function. The chemical potential is derived from the equation for the particle number on the holographic screen \cite{Pazy12}
\begin{equation}
N_{par}={\beta A m \over 2 \pi {\hbar}^2}\int_0^{\infty} { d\epsilon \over \exp{[(\epsilon -\mu)/T}]+1},
\label{eq:number}
\end{equation}
where $\beta=2s+1$, $s$ is the spin of the particle, $m$, is its mass and in two dimensions $N_{par}=N/2$. For two dimensions it also happens, that  the thermal energy, and thus  $\tilde{\mu}$ for fermions and bosons is the same, since the specific heat of a two dimensional gas is identical for fermionic and bosonic degrees of freedom \cite{May64}. 

The interpolating function $\tilde{\mu}$ has two asymptotic properties, which we will consider;  for large temperatures, it is clear that $\tilde{\mu}$ is equal to one since both Fermi and Bose distributions go to the Maxwell distribution at high temperatures; in this limit, the equipartition relation is restored. At low temperatures the thermal energy of a gas of fermions or bosons is proportional to the temperature squared over some energy scale, $T_0$.

The connection of the above quantum statistical considerations with gravity is realized by introducing the Unruh, temperature acceleration relation \cite{Unruh},
\begin{equation}
\label{eq:Unruh}
T={1 \over 2\pi}{\hbar g \over c},
\end{equation}
where $g$ is the acceleration, into Eq. (\ref{eq:energy});
as well as relation (\ref{eq:Area}), combined with the Einstein, energy mass equivalence $E=Mc^2$, obtaining 
\begin{equation}
\label{eq:gen_force}
{GM \over R^2}=\tilde{\mu} g,
\end{equation}
where we have used, $A=4 \pi R^2$. If one considers $\tilde{\mu}$, obtained from quantum statistical considerations, to be the MOND interpolating function; the above equation (\ref{eq:gen_force}) is exactly the MOND gravitational equation, thus QSMEG can be identified with MOND.  $\tilde{\mu}$ the MOND interpolating function, is defined in terms of QSMEG, Eq. ({\ref{eq:energy}), as the ratio of the gravitational energy to thermal energy given by the equipartition value where each degree of freedom carries $ T/2$. 

We know from statistical mechanics that the asymptotic values of $\tilde{\mu}$ for 
high temperatures/accelerations, $\tilde{\mu}=1$,  which in terms of MOND corresponds to Newtonian gravity as obtained in Eq.  (\ref{eq:gen_force}). Again from statistical mechanics we that in the low temperatures/accelerations limit $\tilde{\mu} \propto T^2/T_0$. On the other hand MOND  for low accelerations gives the following equation of motion
\begin{equation}
\label{eq:MOND}
{GM \over R^2}= {g^2 \over a_0}.
\end{equation}
By identifying the MOND interpolating function with $\tilde{\mu}$ obtained in the QSMEG formulation we obtain that  $a_0$, the MOND acceleration scale,  is just the temperature $T_0$, transformed into acceleration via the Unruh relation (\ref{eq:Unruh}). Hence we can associate, by means of QSMEG, the MOND characteristic acceleration scale, $a_0$ to a typical two-dimensional energy scale proportional to the Fermi energy for fermionic dgrees of freedom on the holographic screen, or its analog for the bosonic case. It should be noted that QSMEG does not only lead to MOND, it gives the physical basis for MOND in terms of quantum statistics, as such whereas in MOND the interpolating function, $\tilde{\mu}$, is not defined and is just fixed phenomenologically, QSMEG allows one to calculate $\tilde{\mu}$ analytically, Eq. (\ref{eq:mu_analytical}).

Based on Eq. (\ref{eq:energy}) we now rederive two alternative expressions for $\tilde{\mu}$ obtained in \cite{Pazy13}. Associating, as in Ref. \cite{Padmanabhan12}, the number of bulk degrees of freedom with $N_{bulk}=E/(T/2)$  and inserting this into Eq. (\ref{eq:energy}), we obtain an expression for $\tilde{\mu}$ based on the ratio of the number of bulk to surface degrees of freedom on the holographic screen,
\begin{equation}
\label{eq:degrees_of_freedom}
\tilde{\mu}({ T\over T_0})={N_{bulk}\over N}.
\end{equation}
Another expression for $\tilde{\mu}$, is obtained starting from the thermodynamic  relation for the entropy, $S$ of the holographic screen, worked out in Ref. \cite{Padmanabhan04}, $S=E/2T$, inserting it into (\ref{eq:energy}) we obtain
\begin{equation}
\label{eq:mu_alter}
S={\tilde{\mu} N \over 4}.
\end{equation}
Then, using Eq. (\ref{eq:Area}) and the expression for the Bekenstein-Hawking entropy \cite{Bekenstein,Hawking},
\begin{equation}
\label{eq:BH_ent}
S_{BH}={Ac^3 \over 4G \hbar},
\end{equation}
we identify yet another expression for the MOND interpolating function
\begin{equation}
\label{eq:mu_ent_ratio}
\tilde{\mu} ({T \over T_0})={S \over S_{BH}}.
\end{equation}

\subsection{The universe's acceleration in terms of QSMEG}
\label{sec:dark_ene}

In this section we show that interpreting  gravity in terms of QSMEG is not limited to formulating MOND, based on the formalism of QSMEG it is also possible to obtain the right scale for the cosmological constant and explain the universe's acceleration in terms of an entropic force. We start by briefly describing the cosmological constant problem, then we will recount a solution to this problem, based on the entropy associated with the horizon \cite{Smoot}; consequently showing how similar results, up to some drawbacks, are obtained through QSMEG.

In 1998, based on Supernovae Type 1A data, it was first reported by two groups, that the universe is accelerating \cite{Reiss,Perlmutter}. The acceleration can be can be realized via the Friedmann equation, which describes the evolution of a homogeneous and isotropic universe 
\begin{equation}
\label{eq:Friedman}
H^2={8\pi G  \over 3} \rho,
\end{equation}
where $\rho$ is the energy density source, which drives the expansion of the universe and $H$ is the Hubble parameter, defined in terms of the cosmic scale factor $a(t)$
\begin{equation}
\label{eq:expansion}
H(t)=  {\dot{a} \over a} ,
\end{equation}
where the dot is the time derivative. The accelerated expansion is attained by considering a dark energy (DE) term as part of $\rho$. To justify this term one might try to identify the cosmological constant, $\Lambda$, with the vacuum energy, which automatically leads to the infamous cosmological constant problem \cite{Padmanabhan03}. 

In Ref. \cite{Smoot} an alternative to the DE explanation, for the observed acceleration of the universe, was presented. The basic idea being that the acceleration of the universe is entropic and arises due to the entropy associated with the horizon. Given a horizon temperature 
\begin{equation}
\label{eq:temperature}
T_H=\hbar {H \over 2\pi}.
\end{equation}
An acceleration directly results from the associated horizon temperature through the Unruh relation
\begin{equation}
\label{eq:acc_cosmo}
a_H = {2 \pi c T_H \over \hbar}.
\end{equation}

To find the source of the force responsible for this acceleration and the negative pressure, Ref.  \cite{Smoot} established a connection between the entropy on the horizon and the energy associated with the acceleration. This connection attributes the negative tension, driving the acceleration, to the entropic tension due the entropy of the horizon; rather than resulting from the negative pressure of the DE. We first review the arguments in \cite{Smoot} subsequently we will consider their relevance in the frame of QSMEG.

By associating an entropy with the Hubble horizon, an entropic force  can be obtained
\begin{equation}
\label{eq:force}
F_r=-(dE/dr).
\end{equation}
The associated entropy is given by 
\begin{equation}
\label{eq:entropy_Hor}
S_H={c^3 \over G\hbar} \pi R_H^2,
\end{equation}
where $R_H=c/H$ is the Hubble radius. This leads to the following  entropic force
\begin{equation}
\label{eq:entropic_force}
-T_H{dS \over dr}=-{c^4 \over G}.
\end{equation}
The corresponding pressure $P=(F_r/A)$ is given by
\begin{equation}
\label{eq:pressure}
P=-{c^2 H^2 \over 4 \pi G}.
\end{equation}
This negative pressure in terms of the critical energy density $\rho_{crit} \equiv 3H^2/8\pi G$ is
just 
\begin{equation}
\label{eq:pres_crit}
P=-{2 \over 3} \rho_{crit}c^2.
\end{equation}
This obtained value is close to the current measured negative pressure associated with the cosmological constant. The above derivation, Eqs. (\ref{eq:entropy_Hor}) to (\ref {eq:pres_crit}), of the negative pressure, presented in \cite{Smoot}, is general and relies solely on thermodynamics.  We will present below a different derivation of the negative pressure, based on the energy associated with the holographic screen, which is referred to as surface energy \cite{Pazy13}. 

We start by a short review the derivation \cite{Verlinde_priv}, presented in \cite{Pazy13}, of the surface energy in the deep MOND limit, i.e., the limit  of low accelerations.
For a rotating stellar object with mass $M$ and velocity $v$ in the deep MOND limit the dynamics are
\begin{equation}
\label{eq:DML}
{\left ({v^2 \over R} \right )}^2={G M \over R^2}a_0.
\end{equation}
To calculate the gravitational energy we introduce  $\mid \nabla \Phi \mid^2={G M a_0 /R^2}$, and  calculate the gravitational energy associated with the gravitational field
$E_G(R)=(1/ 8 \pi G) \int dV \mid \nabla \Phi_M \mid ^2$
obtaining
\begin{equation}
\label{eq:Grav_gd_ene}
E_G(R)= {M R  \over 2} a_0.
\end{equation}
The above Eq. (\ref{eq:Grav_gd_ene}), was shown in \cite{Pazy13} to correspond the nonthermal surface energy. It is important to realize that in QSMEG there are two types of energy, the energy of the vaccum related to the ground state energy of the degrees of freedom on the holographic screen and the gravitational energy, which corresponds to thermal excitations. The vaccum energy is proportional to $a_0$ and in the low temperature/acceleration regime the thermal energy is proportional to $1/a_0$ since the number of thermal excitations is proportional to, $T/T_0$. 
The energy calculated in, Eq. (\ref{eq:Grav_gd_ene}) is responsible for the gravitational energy,  which is not  related to thermal excitations. To calculate the pressure associated with Eq. (\ref {eq:Grav_gd_ene}) we employ Eq. (\ref{eq:force}) and divide by the area of the horizon
\begin{equation}
\label{eq:pressure_MOND}
P= -{M  H^2 \over 8 \pi c^2} a_0.
\end{equation}
We insert the expression for the mass by integrating Eq. (\ref{eq:Friedman} ) over the volume, obtaining 
\begin{equation}
\label{eq:}
P= -{c  H \over 16 \pi G} a_0.
\end{equation} 
Inserting the value for the MOND acceleration from Eq. (\ref{eq:cos_a0})
\begin{equation}
\label{eq: pressure_res}
P=-{ c^2 H H_0 \over 32 \pi^2 G}.
\end{equation}

The above result, (\ref{eq: pressure_res}) has to be taken with a grain of salt, first  if one considers $H=H_0$ it differs from the pressure found in Eq. (\ref{eq:pressure}) by a factor of $1/8 \pi$. Nonetheless some of this factor can be explained away considering that  a factor of $2\pi $ of the difference results from the difference between the relation between $a_0$ and $H_0$ as expressed in Eq. (\ref{eq:cos_a0}), to  the relation between $a_H$ and $T_H$ presented in Eqs. (\ref{eq:temperature}), (\ref{eq:acc_cosmo}). However, the more serious pitfall in obtaining the result is the use of the Friedmann equation (\ref{eq:Friedman}) in it's derivation. As we shall see in the next sections the Friedmann equation can not be applied to a MOND like cosmological theory.

\section{Complications and a solution regarding the naive cosmological extension of MOND}
\label{sec:consideration}

MOND is very successful at  modeling galactic rotation curves; however, MOND is a  phenomenological model designed to explain observational data on galactic scales without the need to introduce nonbaryonic DM. In this section we consider attempts to construct a cosmological MOND extension and determine some of the constraints imposed on such a theory through the QSMEG formulation.

\subsection{Naive extensions of MOND to cosmology}
\label{sec:naive}
According to QSMEG the equipartition relation for the degrees of freedom on the horizon needs to be replaced with Eq. (\ref{eq:energy}). In Ref. \cite{Pazy13} it was shown that this  leads in de Sitter space to the following modified Friedmann equation
\begin{equation}
\label{eq:ModdeSitter}
H^2={8 \pi L_p^2 \rho \over 3 \tilde{\mu} (T/T_\Lambda)},
\end{equation}
where, $T_\Lambda=(\Lambda/3)^{1/3}/2\pi$, is some background temperature, $L_p^2 = G \hbar /c^3$ and $T$, is the temperature associated with an acceleration, $g$, according to the Unruh relation (\ref{eq:Unruh}). A  similar more general result was obtained \cite{Zhang} by applying MOND to cosmology for a homogeneous and isotropic universe whereas the low temperature limit  case, corresponding to deep MOND limit, was described in \cite{Kiselev12}. A kin result obtained in a different method was described in \cite{Neto} and previously in \cite{Sanders}. At first sight such an expression seems reasonable especially since in the high acceleration limit $\tilde{\mu}\rightarrow 1$ and the Friedmann equation is obtained. Nevertheless it was noted \cite{Kiselev12} that the above expression (\ref{eq:ModdeSitter}) does not agree with the astronomical observation that we live in a  homogenous and isotropic universe.

In the cosmology theory for a homogenous and isotropic universe the acceleration $g$ depends on the scale factor $a(t)$ as do the velocity $\vec{v}$ and the position $\vec{x}$
\begin{equation}
\vec{x}=a(t)\vec{r}, \; \; \vec{v}=\dot{a}(t)\vec{r}, \; \; \vec{g}=\ddot{a}(t)\vec{r}.
\end{equation}
where, $\vec{r}$, is the co-moving coordinate. Introducing in these terms  a straightforward application of MOND to cosmology leads, as was stated in \cite{Kiselev12}, in the deep MOND limit to the following equation
\begin{equation}
\label{eq:cosmo_mond}
{\ddot{a} \over a} | \ddot{a} |=-{4 \pi G \over 3} (\rho+3 p) {a_0 \over r},
\end{equation} 
where $p$ is the pressure defined by the equation of state for the density $\rho$ and $r=|\vec{r}|$. Equation (\ref{eq:cosmo_mond}), as explicitly stated in \cite{Kiselev12}, is inconsistent with the astronomical observations that the matter in the universe is homogeneous and isotropic on large scales. The reason for the inconsistency is its explicit dependence on the co-moving coordinate, $\vec{r}$. 

\subsection{Modifying the MOND acceleration scale in the cosmological extension of MOND}
\label{sec:mod}
It was suggested in 
 \cite{Kiselev12} that  in the cosmological extension of MOND the typical acceleration scale should be modified to
 \begin{equation}
 \label{eq:acosmological}
 a_0 \rightarrow a_c=g_0|\vec{x}|.
 \end{equation}
 This directly leads to an evolution equation that is consistent with a homogeneous and isotropic universe and as expected does not depend on the co-moving coordinate. In the deep MOND limit corresponding to $|\ddot{a}/a| \ll g_0$, one obtains \cite{Kiselev12},
 \begin{equation}
 \label{eq:DML_cosmologic}
 {\ddot{a} \over a}{|\ddot{a}| \over a}=- {4 \pi G \over 3} (\rho+3p)g_0.
 \end{equation}
 
 Furthermore $g_0$ is parameterized as
 \begin{equation}
\label{eq:parametrization}
 g_0=K_0H_0^2,
 \end{equation}
where $K_0$ is a constant and from astronomical observations $K_0 \sim 10$. The consistency of this suggestion, modifying the MOND acceleration scale according to Eq. (\ref{eq:acosmological}), is then checked with respect to current observational data. In \cite{Kiselev12}, it was shown that using the above cosmological MOND acceleration scale for, $a_c$, with the parameterization (\ref{eq:parametrization}), is consistent with current astronomical observations. Accordingly, there is no need after the modification for any DM to explain the dependence of the brightness of Ia-supernovae data on the redshift. In Ref. \cite{Kiselev} it was shown that the above cosmological extrapolation of MOND, Eqs. (\ref{eq:acosmological}),(\ref{eq:DML_cosmologic}) with the parameterization (\ref{eq:parametrization}) can also successfully reproduce the main features of the CMBR multiple spectrum anisotropy. 

In the next section we will consider what according to QSMEG are some of the constraints on way that MOND can be cosmologically extended, specifically with regards to the typical acceleration scale $a_0$.
 
\subsection{Scaling arguments due to QSMEG regarding the cosmological extension of MOND}
\label{sec:limits}

In Ref. \cite{Kiselev12} the form of the MOND acceleration scale $a_0$, in the cosmological extension of MOND, was suggested to be given by Eq. (\ref{eq:acosmological}). In this way  MOND (without additional fields) can be extended consistently with observational data showing we live in  a homogeneous and isotropic universe. In this section we give two thermodynamic scaling arguments showing why the MOND typical acceleration scale $a_0$, should scale as in Eq. (\ref{eq:acosmological}). These arguments are both based on the formulation of  MOND in terms of QSMEG. The first argument is due to entropy considerations, the second due to the connection derived in QSMEG between $a_0$ and the Fermi energy of the holographic screen degrees of freedom. 

According to QSMEG one of the possible definitions of the MOND interpolating function is given by Eq. (\ref{eq:mu_ent_ratio}), where the interpolating function is given as a ratio between $S$, which  is the two-dimensional entropy of the holographic screen to the $S_{BH}$, which is the Bekenstein-Hawking entropy \cite{Bekenstein, Hawking}. This definition was shown \cite{Pazy13} to hold for a  cosmological extension of MOND to de Sitter space. As the universe expands, the entropy related to its horizon, grows in proportion to its area, $S \propto A$; however, also the Bekenstein-Hawking entropy Eq. (\ref{eq:BH_ent}) is proportional to the horizon area. As such the MOND interpolation function should stay constant throughout the universe's expansion. Since $\tilde{\mu}[\ddot{a}(t) r/ a_0]$ is a function of the ratio of the acceleration to the MOND acceleration scale, and since $\tilde{\mu}$ is fixed during the universe's expansion, it is clear that the typical MOND acceleration scale should scale as  $a_c$, in Eq.(\ref{eq:acosmological})

The second scaling argument is based on the connection between $a_0$ and the Fermi energy. In Ref. \cite{Pazy12} it was shown that the MOND acceleration scale $a_0$ can be the identified with, 
a typical energy scale, which divides between the quantum and the classical regimes, which is proportional to the Fermi energy, $E_F$, in the fermionic case,
\begin{equation}
\label{eq:Fermi_ene}
a_0=({12 c \over \hbar \pi})E_F.
\end{equation}
Given that in two dimensions the Fermi energy is linear in the particle number, $a_0$ should scale like the number of degrees of freedom on the horizon or rather the effective number of degrees of freedom. In Ref. \cite{Padmanabhan12} the number of effective degrees of freedom, $N_{eff} \equiv (v^2/2c^2)N_{surf}$, created by a mass $M$ on a holographic sphere of radius $R$ was calculated to be 
\begin{equation}
\label{eq:N_eff}
N_{eff}=2 \pi {M R c \over \hbar}.
\end{equation}
Since $a_0$ is proportional to $E_F$, which in turn is proportional to $N_{eff}$ we find that according to QSMEG, $a_0$ should scale with $R$, leading again to expression (\ref{eq:acosmological}).

\section{ Complications regarding a more general cosmological extension of MOND and possible solutions}
\label{sec:complications}

In the previous section thermodynamic scaling arguments where given in order to determine how according to the QSMEG formulation, MOND should be cosmologically extended. The question regarding the constraints on cosmologically extending MOND and more specifically regarding the scaling of the MOND acceleration parameter $a_0$ where studied from a different aspect in \cite{ Starkman,Lue04PRL}. 
In these references by requiring that the underlying gravity theory for MOND be compatible with Birkhoff's law and an homogenous universe,  a generalized modified Friedmann equation for MOND is obtained. Then by the requirement that $a_0$ distinguish between MOND like and Einstein dynamics, the functional form of $a_0$ is determined. According to the calculations presented  in \cite{ Starkman,Lue04PRL}, Birkhoff's theorem dictates a weak dependence of $a_0$ on the source mass $a_0 \sim M^{1/3}$. This dependence is then shown to cause a suppression of structure growth, which both papers view as significant difficulty for a cosmological extension of MOND. In the section below, we briefly describe the arguments put forth in \cite{ Starkman,Lue04PRL} and then subsequently show how the problem is resolved if one considers that on a cosmological scale the typical acceleration scale is given by, $a_c=g_0|x|$.

\subsection{Constraints on the form of the MOND acceleration scale from Birkhoff's theorm}
\label{sec:constraints}

In \cite{ Starkman,Lue04PRL} an alternative Friedmann equation is considered, as the most general form, which an extension of MOND would also follow,
\begin{equation}
\label{eq:Fried_alt}
{\left ({\dot{a} \over a} \right )}^2=H_0^2 g(x),
\end{equation}
where $x=8 \pi G \rho /3H_0^2$, is a dimensionless parameter.
The function $g(x)$ is determined by the scale factor, $a(t)$ under the following constraints; in requiring that a cosmological extension of MOND be compatible both with Eq. (\ref{eq:Fried_alt}) and with Birkhoff's theorem the functional form of $g(x)$ is determined. Requiring that Eq. (\ref{eq:Fried_alt}) be compatible with MOND defines the MOND acceleration scale as  \cite{ Starkman,Lue04PRL}
\begin{equation}
\label{eq:MOND_starkman}
a_0=H_0[9 \beta^2(r_gH_0)^{1/3}],
\end{equation}
where $r_g=2GM$, is the Schwarzschild radius of a matter source $M$, and $\beta$ is a constant parameter, chosen \cite{ Starkman,Lue04PRL} to be $\beta \approx 15$ so as to obtain $a_0 \approx c H_0 /2 \pi$. 

The expression for $a_0$, Eq. (\ref{eq:MOND_starkman}) was thus obtained on very general arguments; however, as shown in \cite{ Starkman,Lue04PRL} this result  is problematic for MOND. The weak dependence of $a_0$ upon the mass, $a_0 \sim M^{1/3}$, results in the suppression of structure growth such that a cosmological MOND theory would not be able to account for the observed structure formation in the universe. In \cite{ Starkman}, Starkman states that "in real MOND"  the problem does not exist since $a_0$ does not scale with $M$; however, one cannot accept this as the solution to the problem since this dependence is a result of Birkhoff's theorem, and to relinquish it would have dire consequences. In the section below we show this problem is resolved if one assumes the validity of Eq. (\ref{eq:acosmological}).

Actually the above dependence, expressed in Eq.  (\ref{eq:MOND_starkman}) of $a_0$ on the mass, was also noted in \cite{Kiselev12}; nonetheless, the authors found it to be compatible with Eq. (\ref{eq:acosmological}). The argument presented in \cite{Kiselev12}, was that since $a_0 \sim M^{1/3}$ and since a homogeneous distribution of matter with a fixed density $\rho$ implies $M \sim \rho  |x|^3$ accordingly $a_0 \sim |x|$, as was postulated in Eq. (\ref{eq:acosmological})  for $a_c$. However, it turns out that if the dependence presented in Eq. (\ref{eq:acosmological}) is taken into consideration, once the conditions on $a_0$ are determined, one finds a modified MOND acceleration scale $g_0$, which has no explicit mass dependence.

\subsection{A possible solution to the problem of the mass dependence of the MOND acceleration}
\label{sec:solution}

In Refs. \cite{ Starkman,Lue04PRL} the condition for defining $a_0$ as it appears in Eq. (\ref{eq:MOND_starkman}) was found by the continuity equation for the acceleration given by
\begin{eqnarray}
\label{eq:acceleration}
a =  \left \{ \begin{array}{ll}
-\frac{1} {2}\frac{r_g}{r^2}, &  |a|>a_0,  \\
-\frac{3 \beta}{2}\frac{(r_gH_0)^{2/3}} {r}, & |a|<a_0.
\end{array} 
\hspace*{10cm}
\right.
\end{eqnarray}
The corresponding radius,  for which the above two expressions are equal to one another, is given by
\begin{equation}
\label{eq:cors_rad}
r=r_g/3 \beta (r_g H_0)^{2/3}.
\end{equation}
Now if we require that $a_c=g_0|x|$ one needs to divide the previous result  for $a_0$ in Eq. (\ref{eq:MOND_starkman}), by the value of the above corresponding radius, Eq. (\ref{eq:cors_rad}), hence obtaining a modified result, which does not depend any more on, $r_g$, the Schwarzschild radius of a matter source,
\begin{equation}
\label{eq:mod_a0}
g_0= 27 \beta^3 H_0^2.
\end{equation}
This result corresponds to Eq. (\ref{eq:parametrization}), in which the constant 
 $K_0$ appearing in front of $H_0^2$ was  found to be  from observations  $ \sim10$.
Equation (\ref{eq:mod_a0}) does not have the problematic scaling dependence on the mass. This is essentially the reason why the authors in Ref. \cite{Kiselev12} were able to describe the large scale structure of the universe with a cosmological extrapolation of MOND. The problem of the dependence of $a_0$ on the source mass, which causes suppression of structure growth in the universe, is thus resolved by modifying MOND on the cosmological scales by Eq. (\ref{eq:acosmological}), in which $a_c=g_0|x|$. 

\section{Possible connections between cosmological extended MOND and alternative gravity theories}
\label{sec:connections}

The extension of MOND to cosmology discussed so far, clearly does not correspond to GR. The question remains can  MOND theory be incorporated into some alternative gravity theory; in which MOND would correspond to the classical limit? In this section we investigate this issue, alluding to the connection between MOND and two, alternative gravitational theories, both of which are defined in terms of physical scales, which do not appear in GR. These theories are Conformal gravity (CG) and an alternative gravity theory developed by Grumiller \cite{Grumiller10}, involving a Rindler force, which  we shall refer to it as Rindler force theory (RFT).

Conformal gravity (CG), which was first introduced by Weyl \cite{Weyl}, has been intensively studied and greatly developed over the last 25 or so years by Manneheim (for a recent review see \cite{Mannheim}). The theory was originally advanced by Manneheim as an alternative candidate to
GR since it allows a nontrivial de Sitter geometry solution without a cosmological constant. In CG due to its conformal symmetry the cosmological term has to vanish identically. Moreover it turns out that CG is also able to fit galactic rotations with no additional galaxy dependent parameters. In a recent paper, Mannheim and O'Brien \cite{Mannheim12}, presented a comprehensive fit of 141 galaxies using CG with no need for galactic dark matter. The fit presented in \cite{Mannheim12} was based on an analytical solution of the CG obtained by Mannheim and Kazanas,  which allowed them to obtain the nonrelativstic limit of the theory \cite{Mannheim94}. CG theory is based on the principle of local conformal symmetry, which leads to fourth order equations of motion. Utilizing the underlying conformal symmetry of the theory Mannheim and Kazanas were able to obtain a simple and compact form for the equations of motion defining the static and spherically symmetric case. The solution to this equation of motion, being a fourth order differential equation, is fixed by four integration parameters, which introduce new physical scales. The values of the these integration parameters are obtained empirically by astronomical observation and once they are fixed they cannot be adjusted per galaxy (in contrast to DM fits).
Fitting galactic rotation curves, is then preformed by summing contributions to circular velocities due to local material
within the galaxies $v^2_{loc}$ and also due to cosmological terms
\begin{equation}
\label{eq:rotation}
v_{tot}^2(R)=v_{loc}^2(disk)+{\gamma_0 c^2 R \over 2}-\kappa c^2 R^2,
\end{equation}
where $\gamma_0=3.06 \times 10^{-30}$ cm$^{-1}$ is an integration parameter related to a universal linear potential and  $\kappa=9.54 \times 10^{-54}$ cm$^{-2}$ is a further integration parameter related to a universal quadratic potential. The asymptotic limit for Eq. (\ref{eq:rotation}), 
\begin{equation} 
\label{eq:asymp}
v_{tot}^2(R) = {N^* \beta^* c^2 \over R}+{N^* \gamma^* c^2 R \over 2}+
{\gamma_0 c^2 R \over 2}-\kappa c^2 R^2,
\end{equation}
is defined by further two integration parameters:  $\gamma^*=5.42 \times 10^{-41}$ cm$^{-1}$ and $\beta^*$, which can be simply identified as the Schwarzschild radius of the sun.  $N^*$ being the number of stars within the galaxy. 

It is of interest to compare Eq. (\ref{eq:asymp}) with  regular Newtonian gravity; in doing so it should be noted that in CG, due to the conformal symmetry, Newton's constant $G$, cannot appear, instead we see that it is replaced by $ \beta^* c^2/M_{\odot}$, where $M_{\odot}$ is one solar mass. Hence, the first term on the right in Eq. (\ref{eq:asymp}) is just the Newtonian gravitational potential.  $\kappa$ is relatively small and thus for relatively short distances it can be neglected; nonetheless, it turns out it is imperative to include the $-\kappa c^2 R^2$ term to fit galactic rotation curves \cite{Mannheim12}. According to CG the departure from Newtonian dynamics on galactic scale is systematic and is given by $ (R c^2 / 2)(N^* \gamma^*+ \gamma_0)-\kappa c^2 R^2$. Current observational data seems to fit very well within this prediction \cite{Mannheim12}.

Another alternative gravity theory, RFT, developed by Grumiller, involves a Rindler force . RFT is formulated by statisfying a number of constraints expected from a reasonable theory of gravity. The theory has to be well behaved at large distances, i.e., asymptotically gravity is required to have a diffeomorphism invariance and is spherical symmetric, which effectively reduces it to a two-dimensional theory. Gravity is also expected to have, power counting renormalizability and local validity of Newton's law, as well as cosmic censorship. Fulfilling all these requirements leads to an alternative gravity theory incorporating an effective potential, which is composed of a cosmic potential defined by a cosmological constant, and a Rindler force proportional to $a_{R}$ the Rindler acceleration. Circular velocities in terms of RFT are given on galactic scales by \cite{Grumiller10}
\begin{equation}
\label{eq:vel_RFT}
v(R) \approx \sqrt{{ G M(R) \over R}+a_{R}R},
\end{equation}
where the Rindler acceleration, $a_R$, can be a function of $R$. For a constant $a_R$ the corrections to the Newtonian potential, as in CG, are linear in $R$ on galactic scales. It should be noted that the restrictions stated above, which were  employed in constructing RTF, do not exclude  the possibility of a cosmological constant term and thus the RTF effective potential can also include a $-\Lambda R^2/2$ term, where $\Lambda$ is some cosmological constant.

\subsection{Some numerical connections  between MOND to CG and RFT}
\label{sec:alter}

The two alternative gravity theories, CG and RTF, briefly described above have additional physical scales, which do not appear in GR. Comparing one of  the two cosmological related constants in CG, $\gamma_0$, to the acceleration scale in MOND, $a_0$ , we obtain
 $\gamma_0 c^2 \sim a_0$ and since, $\gamma_0$, is cosmologically  related it is not surprising to find that  $\gamma_0 c \sim H_0 \sim a_0/ c$. The second cosmologically related parameter in CG is  $\kappa$; taking the  value of $g_0$, according to Eq. (\ref{eq:parametrization}),
 where $K_0 \sim 10$ we obtain $g_0 \sim \kappa c^2$. In RTF the physical scale of the Rindler acceleration depends on the system at hand. In Ref. \cite{Grumiller10} working in units of $c=\hbar=G=1$, the typical scale for the Rindler accelerations for dwarf galaxies, as well as for our galaxy, is $a \approx 10^{-62}$ whereas in these units one also obtains that  $a_0 \approx 10^{-62}$.

In CG and RFT as well as the cosmological extension of MOND parameters are fixed by observations so it might not be of surprise that they turn out to be similar in scale, but the important thing to note is that the galactic scale is related to the cosmological scale $H_0$.  In the next section we will demonstrate that the similarities between MOND, CG and RFT, are not limited to the numerical values of the parameters appearing in these theories, rather these theories exhibit the same physical behavior in limiting cases.

\subsection{Connections between MOND with CG and RTF in terms of galactic rotation curves}
\label{sec:conformal}

The alternative gravity theories described above, CG and RTF  as well as MOND predict that there should be regularities in the observational data on galactic and extragalactic scales. Observationally one of the most distinct correlations noted in galactic data is the TF relation \cite{TF}. The baryonic TF relation is an empirical  relationship between the circular velocities of galaxies $v_c$ and the baryonic mass content, $M_b$, of these galaxies, as estimated from the luminosity
\begin{equation}
\label{eq:TF}
M_b \propto v_c^4.
\end{equation}
This relation was shown to hold for over five decades in stellar mass and for velocities ranging from 
30 to 300 km s$^{-1}$, where small adjustments need to be made in order to obtain the mass from the observed luminosity for faint galaxies, due to their high concentration of gas \cite{McGaugh00}. Such regularities are very difficult to explain in terms of DM (for a proposed explanation see \cite{Steinmetz99}) and in general call for fine tuning of DM parameters. 

The above relation (\ref{eq:TF})  is consistent with a constant acceleration scale $a=v_f^4/(GM_b)$,
where $v_f$ is the velocity corresponding to the flat part of the rotation curve. According to  MOND in the deep MOND limit,
\begin{equation}
\label{eq:TF_equation}
v_f^4=G M_b a_0.
\end{equation}
This correspondence occurs since, MOND was essentially constructed such that its acceleration scale can  be identified with the Tully-Fisher acceleration.

Below we demonstrate how the baryonic TF relation is obtained in CG and RTF, using this to connect MOND to these theories. One should keep in mind that the theoretical motivations for constructing CG and RFT were completely different and the fact that it can be shown that they obey the TF relation should be considered as a confirmed prediction of these theories. The observation that MOND can be considered as a limit case of CG in some special case was first made by Mannheim and most of the results we present below actually appear as a footnote in \cite{Mannheim06}. However, Mannheim does stress some distinctions between MOND and the nonrelativstic limit of CG, which we shall briefly elaborate on.

We start by considering Eq. (\ref{eq:asymp}) for CG at the maximum of the velocity with respect to distance. To do so we consider the radius, $R_m$, for which the derivative of the right hand side of Eq. (\ref{eq:asymp}) with respect to $R$  is equal to zero, as the radius defining $v_f$. In taking the derivative, the term corresponding to the cosmological parameter $\kappa$ is neglected since it is relatively small on galactic and extragalactic scales, thus obtaining 
\begin{equation}
\label{eq:TF_CG}
{N^* \beta^*  \over R_m^2}={N^* \gamma^* +\gamma_0  \over 2}.
\end{equation}
We consider two typical cases: (1) $N^* \gamma^* \ll \gamma_0$, which corresponds to dwarf galaxies, (2) $N^* \gamma^* \sim \gamma_0$, corresponding to spiral galaxies. 

For case (1), we insert the relevant value obtained for $R_m$ from Eq. (\ref{eq:TF_CG}), into the expression  for $v_f^4$ obtained by squaring Eq. (\ref{eq:asymp}), 
\begin{equation}
\label{eq:case1}
{v_f}^4= 2 c^4 N^* \beta^* \gamma_0.
\end{equation}
Recalling that $\beta^*$ is the Schwarzschild radius of the sun and that $N^*$ is the number of stars within the galaxy, we obtain that Eq. (\ref{eq:case1}) is equivalent to Eq. (\ref{eq:TF_equation}) with $a_0=2 c^2 \gamma_0$. This dependence leads to a universal gravitational law, in which there is departure from Newtonian gravity, which depends on the single acceleration scale $a_0 \sim 2 c^2 \gamma_0$. There is a major distinction in obtaining the TF relation through MOND or through CG. Whereas in MOND the acceleration scale is a phenomenological scale, in CG the $\gamma_0$ related potential is of cosmological origin, it results from the gravitational force an orbiting object feels due to expansion of the universe \cite{remark_Gauss}. 

However, in CG there is also an $N^*\gamma^*$ term, which has so far been neglected. When considering spiral galaxies, case (2), the $N^* \gamma^*$ term can no longer be neglected. This term leads to a further distinction between MOND and the nonrelativistic limit of CG. Whereas the gravitational force in MOND is the Newtonian gravitation, in its CG equivalent there 
also is an added force proportional to $N^*\gamma^*$. Due to this extra term it first seems that one cannot expect a TF relation to hold in spiral galaxies; yet, according to \cite{Mannheim06} the reason for the universal behavior in these galaxies, is that they obey a Freeman law of constant surface brightness, $\Sigma^F_0$. The constant surface brightness condition gives a common value for $N^*/2 \pi R_0^2$ in these galaxies, where $R_0$ is the typical length scale defining the distribution of stars within these thin disk galaxies $\Sigma(R)=\Sigma_0^F e^{-R/R_0}$.

It turns out that for these galaxies the luminosity is controlled almost entirely by the luminous Newtonian contribution 
\begin{equation}
{v_{lum}^2\over c^2} \sim 0.8 \pi \Sigma_0^F\beta^*R_0.
\end{equation}
The correction to this value from $\gamma^*n^*+\gamma_0$ being around 15$\%$; nonetheless, it turns out that in CG for Freeman galaxies the universality of $\gamma_0$ and $\Sigma^F_0$ are connected through
\begin{equation}
\label{eq:Freeman}
\gamma^*N^*+\gamma_0=0.12\pi\Sigma_0^F\beta^*.
\end{equation}

The fact that in CG, $\gamma_0$, is of cosmological origin and it is related to $\Sigma_0^F$ through 
(\ref{eq:Freeman}) hints that  $\Sigma_0^F$ is of cosmological origin and its value is related to galaxy formation restrictions, as such it might explain the Freeman law of constant surface brightness. It should be noted that in CG unlike Newtonian gravity there is a maximal radius for galaxies.

Using similar considerations  one can obtain the TF relation in terms of RTF. By first taking the derivative of the right hand side of  Eq. (\ref{eq:vel_RFT}) and equating it to zero, one obtains the following equation for $R_m$ in terms of
RTF
\begin{equation}
\label{eq:RTF}
{G M \over R_m^2} =a_{R}.
\end{equation}
Inserting the above expression for $R_m$ into the expression for ${v_f}^4$ obtained by squaring Eq. (\ref{eq:vel_RFT}), the following TF relation is obtained
\begin{equation}
\label{eq:TF_RTF}
{v_f}^4=4  G M a_R.
\end{equation}
Comparing this equation with Eq. (\ref{eq:TF_equation}) one obtains, $a_0=4 a_R$, which fits very well with the observational data \cite{Lin13}, from which a value of $a_R \approx 0.3 \times 10^{-10} $ m s$^{-2}$ was obtained. It should be mentioned that though Ref. \cite{Lin13} seems to get very good fits with $a_R$ corresponding to $a_0/4$ these result have been recently criticized in Ref. \cite{Mastache13}, in which it was claimed that the good fit is the result of the limited amount of galaxies considered by the authors and once a larger sample is considered, the good correspondence of RTF to galactic rotation curves no longer holds.

The controversy with regards to the quality of the RTF fits to the observational data is actually related to the historical efforts to fit galactic rotation curves with CG, culminating with the recent fit of 141 galaxies by Mannheim and O'Brien \cite{Mannheim12}. Originally Mannheim attempted to fit the galactic rotation curves with the potential corresponding to $v_{loc}^2(disk)$, without the $\gamma_0$ linear term, allowing $\gamma^*$ to vary
\cite{Mannheim93}. Such a fit seems very intuitive since one assumes due to Gauss's theorem that a rotating body feels only the interior forces of stars inside a galaxy and not forces due to the rest of the masses in the universe; all the same, this does not work, since Gauss's theorem does not apply to CG in the classical limit since it involves solving a fourth order differential equation. Mannheim noted that the results could be made to fit much better provided the quantity $N^*\gamma^*$ was universal, which indicted that there should be another linear term in the potential, which depends on the mass sources outside the specific galaxy. Fitting the observations with a linear potential allowing the coefficient to vary is exactly like fitting the data with RFT and it turns out the coefficient comes out universal. Mannheim then tried to fit the observational data ignoring the local linear potential, i.e, the $N^* \gamma^*$ term, which is essentially the same fit being done by RFT, his results were not on the spot, with maximal errors of around 10$\%$ of each data point from the fit \cite{Mannheim95}. This fit is yet again the same fit applied for RFT. As it turned out one needs only to include the $N^* \gamma^*$ term and the fit for the original 11 galaxy survey \cite{Mannheim95}, available at the time, was in excellent agreement with the data. However, when one considers more galaxies this is not enough and one also needs to include the potential $-\kappa c^2 R^2/2$, induced by inhomogeneities in the cosmic background, this was done by Mannheim and O'Brien \cite{Obrien11} to fit a sample of 110 spiral galaxies and recently the comprehensive set of 141 \cite{Mannheim12}. Looking at the galaxies, which were  fitted to a good agreement by RFT \cite{Lin13}, comparing to the 11 galaxies Mannheim originally used ,one finds that the 6 of the 8 galaxies used  by Ref. \cite{Lin13} to fit RFT, were also used in 11 galaxy fit by Mannheim \cite{Mannheim95}
who reported a maximum 10$\%$ error of each data point. These galaxies where: NGC2403, NGC2841, NGC2903, NGC3198, NGC7331 and DDO154. These galaxies were also used in Ref.
 \cite{Mastache13}, which claimed that the fit was good but once you added more galaxies it no longer worked. Similar to the findings made by Mannheim years ago.
 
\section{ Conclusions}
\label{sec:conclusions}
Modifying the entropic formulation of gravity by considering the quantum statistics of the degrees of freedom on the holographic screen, is the basis of the QSMEG formulation of gravity, from which one can in turn obtain MOND. In this paper we have hinted at how one can obtain through QSMEG the accelerated expansion of the universe, via entropic considerations. The difficulties in obtaining a cosmological extension of MOND were specified and it was shown how they could be avoided using scaling considerations, obtained via thermodynamic arguments, based on QSMEG. 

In addition the similarities between MOND and CG and RFT were specified and the two alternative gravity theories, were considered as possible cosmological extensions of MOND. Since MOND can be obtained in terms of QSMEG, which is defined in terms of the quantum statistics of the degrees of freedom on the holographic screen, it is of interest to define CG and RFT in terms of an entropic theory defined on a holographic screen. This subject is left for future research.

\section{Acknowledgments}
The author would like to thank Philip Mannheim for very informative discussions and the physics department of the University of Connecticut for its hospitality.


\begin{thebibliography}{99}
\bibitem{remark_Gauss}
 Being a fourth order derivative theory one cannot not evoke Gauss's law for an orbiting object in CG. In other words the gravitational force in CG is not due only to the matter enclosed within the orbit, it is also due to a force imposed by the rest of the universe.
\end{thebibliography}
\end{document}